# Growth-dependent Interlayer Chiral Exchange and Field-free Switching

Yu-Hao Huang[1], Chao-Chung Huang[1], Wei-Bang Liao[1], Tian-Yue Chen[1], and Chi-Feng Pai[1,2]*

[1]*Department of Materials Science and Engineering, National Taiwan University, Taipei 10617, Taiwan*

[2]*Center of Atomic Initiative for New Materials, National Taiwan University, Taipei 10617, Taiwan*

Interfacial Dzyaloshinskii-Moriya interaction (DMI) has long been observed in normal metal/ferromagnetic multilayers, enabling the formation of chiral domain walls, skyrmions and other 2D antisymmetric spin textures confined within a single ferromagnetic layer, while more recent works on interlayer DMI reveal new pathways in realizing novel chiral 3D spin textures between two separate layers. Here, we report on interlayer DMI between two orthogonally magnetized ferromagnetic layers (CoFeB/Co) mediated by a Pt layer, and confirm the chiral nature of the observed effective field of up to 37 Oe through asymmetric hysteresis loops under in-plane field. We highlight the importance of growth-induced in-plane symmetry breaking, resulting in a sizable interlayer DMI and a universal characteristic vector through wedge deposition of the samples. We further perform deterministic current-driven magnetization switching in the perpendicularly magnetized Co layer utilizing solely the effective field from the interlayer DMI. These results demonstrate interlayer DMI's potential to facilitate deterministic field-free switching in spin memory applications.

* Email: cfpai@ntu.edu.tw

In the field of spintronics, much endeavor has been put into engineering the various couplings between separate magnetic bilayers, which are often the results of either symmetric or anti-symmetric exchange interactions. Symmetric exchange interactions are relatively well researched and documented, directly leading to the existence of ferromagnetic (FM) and antiferromagnetic (AFM) couplings, and important interlayer coupling effects such as the Ruderman-Kittel-Kasuya-Yosida (RKKY) [1-4] interaction. Antisymmetric interaction, or the Dzyaloshinskii-Moriya interaction (DMI) [5,6] on the other hand has gained significant attention lately due to its capability to mediate noncollinear chiral spin configurations. The occurrence of the DMI requires two key factors, first being the existence of inversion symmetry breaking in the material system, second is the requirement of the mediating electrons being subject to strong spin-orbit coupling (SOC) [7,8]. Some prominent examples that take advantage of the DMI include skyrmions [9-11] and Néel domain wall logics [12,13].

Research on the DMI was originally focused on 2D configurations, that is, the two FM atoms described by the three-site Lévy-Fert model [14,15] belong to the same layer. Recently, theoretical calculations [15] have shown the possibility to extend this intralayer phenomenon to the 3D case in which the two FM atoms are from two separated layers, and was experimentally verified shortly after [16-18]. Compared to the 2D case, theoretical calculations and pioneering experimental works unambiguously stated the addition of an in-plane symmetry breaking element is of paramount importance [15,17]. This chiral interlayer coupling effect mediated by the DMI can potentially play a crucial role in future multilayered spintronic devices. Prior to our work, experimental reports utilized either two perpendicularly magnetized layers with FM/AFM coupling [16,17], or employed a ferrimagnetic layer with bulk perpendicular magnetic anisotropy (PMA) alongside a FM layer with in-plane magnetic anisotropy (IMA) [18]. While very recent works demonstrated magnetization switching utilizing the interlayer DMI's effective field, these efforts focused on AFM coupled systems [19,20]. Therefore, by showing the interlayer DMI's capability to mediate current-induced field-free switching in an orthogonally magnetized system, we aim to introduce a new pathway for interlayer DMI's potential usage on more practical magnetic memory devices as well as its correlation to the materials growth approach.

In this work, we demonstrate experimentally the existence of a sizable interlayer DMI coupling between two widely used ferromagnetic layers, Co with PMA and CoFeB with IMA, mediated by a Pt spacer. Pioneering work by Avci *et al*. [18] already demonstrated Pt to be a suitable mediating spacer layer in comparison to other heavy metals (HM), for Pt possesses strong SOC, consistent with the reported HM/FM interfacial DMI scenarios [21-23]. We start by establishing a carefully designed magnetic field sweep procedure, allowing us to accurately quantify effective DMI fields and confirm the chiral nature of the DMI. The importance of in-plane symmetry breaking is then examined by comparing samples with and without symmetry breaking, and subsequently used devices with various orientations to check the universality of the DMI strength and direction on a single substrate. Moreover, we show that current-induced field-free magnetization switching of a PMA layer is achievable through the DMI effective field even with minimal spin torque acting upon the PMA layer, and the switching polarity can be well controlled by manipulating interlayer DMI's characteristic direction.

The main samples used in this work are CoFeB(2)/Pt(2.5)/Co(0.6)/Pt(2.5) (numbers in parentheses are in nanometers) multilayers deposited by high-vacuum confocal magnetron sputtering (base pressure ~$10^{-8}$ Torr and Ar working pressure of $3\times10^{-3}$ Torr) onto Si/SiO$_2$ substrate. Nominal composition of the CoFeB was 40% Co, 40% Fe and 20% B (atomic percentage). Throughout deposition, rotary of the substrates was disabled, with the CoFeB atom flow direction striking substrate normal at α = 25° to produce a wedged multilayer and thus introduce in-plane symmetry breaking (Fig. 1(a)). However, Pt was grown in a normal-incidence configuration with α = 0°. Of the two FM layers, CoFeB has IMA since its thickness far exceeds the spin reorientation transition thickness [24,25] and Co possesses PMA due to strong interfacial anisotropy at the two Co/Pt interfaces [26,27]. For anomalous Hall resistance ($R_H$) and unidirectional magnetoresistance (UMR) measurements [28,29], Hall bar devices with 5-μm width and various rotary angles with respect to the substrate are prepared with standard photolithography processes.

We now investigate the interlayer DMI's effect on a magnetic bilayer system with orthogonal magnetizations. The overall Hamiltonian $\hat{H} = J_H \boldsymbol{M}_1 \cdot \boldsymbol{M}_2 - \boldsymbol{D} \cdot (\boldsymbol{M}_1 \times \boldsymbol{M}_2)$ consists of two parts, first being the conventional symmetric (Heisenberg) exchange term, and the second being the antisymmetric (DMI) term [7,30]. $J_H$ denotes the Heisenberg exchange integral, $\boldsymbol{M}_{1,2}$ denotes the magnetizations of the two magnetic layers, and $D$ is the characteristic DMI vector that governs the strength and direction of the interlayer DMI. Since the two magnetizations form an orthogonal configuration, Heisenberg exchange's contribution to the Hamiltonian is conveniently excluded while the DMI maximized. Due to the degree and direction of symmetry breaking being determined by the sputtering process, $D$ is expected to be fixed and lie within the $xy$ plane due to symmetry constraint [15]. Overall, the system's Hamiltonian is governed by a cross product term $\widehat{H}_{DMI} = -D \cdot (M_1 \times M_2)$. From this Hamiltonian we can take direct analogy from the well-known Zeeman Hamiltonian $\widehat{H}_{Zeeman} = -m \cdot H$, where $m$ is the magnetic moment and $H$ the external field. Thus, macroscopically speaking, interlayer DMI causes the IMA layer to exert an effective field on the PMA layer, and vice versa. If the aforementioned $M_1$ and $M_2$ are chosen to be $M_{PMA}$ and $M_{IMA}$, respectively, then an DMI effective field $H_{DMI} = -D \times M_{IMA}$ is felt by the PMA (Co) layer while the IMA (CoFeB) layer experiences an effective field $H_{DMI} = D \times M_{PMA}$. For example, if the precise direction of $D$ points toward $\varphi_D$=150° direction ($\varphi_D$ denotes the angle between $x$ axis and $D$), then when $M_{IMA}$ is pinned at $\varphi_H = 60°$ direction ($\varphi_H$ being the angle of the applied in-plane field $H_\varphi$), $H_{DMI}$ exerted on $M_{PMA}$ is at maximum and points toward +z direction. Conversely, when $M_{PMA}$ points toward +z, $H_{DMI}$ exerted on $M_{IMA}$ points toward $\varphi_H = 60°$ direction (Fig. 1(a) left).

We devised a field sweep protocol to test the first CoFeB(2)/Pt(2.5)/Co(0.6)/Pt(2.5) wedged-deposited structure patterned into a Hall bar device. An in-plane field $H_\varphi$ is applied to fully align $M_{IMA}$ toward $\varphi_H$. Note that $|H_\varphi|$ must be sizable to avoid the reciprocal $H_{DMI}$ and possible in-plane anisotropy from causing unwanted tilting of $M_{\text{IMA}}$. Subsequently, field sweep is performed in the $\pm z$ direction to capture hysteresis behavior of $M_{\text{PMA}}$ by measuring its $R_H$. Graphical representation of the procedures is demonstrated in Fig. 1(a) lower left panel. Fig. 1(b) shows two hysteresis loops of the Co layer when subject to two different in-plane field angles, namely $\varphi_H = 60°$ and $240°$ with $|\boldsymbol{H}_\varphi| = 300$ Oe. Square $R_H$ hysteresis loops indicate strong PMA character of the Co layer with coercive field $H_c = 25$ Oe. The two hysteresis loops have significantly shifted switching boundaries and we subsequently define the horizonal shift of the loop centers, $H_{\text{offset}}$, by the average of the two switching fields of magnetization's down-to-up and up-to-down transitions. This sizable $H_{\text{offset}}$ is of striking difference with two control samples, namely wedged Pt(2.5)/Co(0.6)/Pt(2.5) and non-wedged CoFeB(2)/Pt(2.5)/Co(0.6)/Pt(2.5), in which PMA hysteresis loops are always symmetrical to the origin with negligible $H_{\text{offset}}$ ($< 2$ Oe) regardless of $\varphi_H$ (Fig. 1(c)). The data of the two control samples also serve as demonstration of the precision of our elaborately calibrated vector magnet, excluding the possibility of artifact related asymmetries from the vector magnet.

Next, following the identical field sweep protocol with a constant $|\boldsymbol{H}_\varphi| = 300$ Oe as $\varphi_H$ rotated from 0 to $360°$, we plot $H_{\text{offset}}$ as a function of $\varphi_H$ as shown in Fig. 1(c). $H_{\text{offset}}$ has a clear sinusoidal dependence on $\varphi_H$ which nicely agrees with the previous discourse in which $M_{\text{IMA}}$ aligns toward $\varphi_H$, leading to $M_{\text{PMA}}$ experiencing a constant effective field with the sinusoidal form $-\boldsymbol{D} \times \boldsymbol{M}_{\text{IMA}}$. The sine fit to the data $H_{\text{offset}} = H_{\text{DMI}} \sin(\varphi_H - \varphi_D)$ suggests an amplitude of $H_{\text{DMI}} = 37$ Oe and $\varphi_D$ of $150°$ for this specific sample. The direction of $\varphi_D$ is decently perpendicular to CoFeB's atom flow direction (Fig. 1(a) upper right diagram), which agrees well with theoretical prediction [17]. The origin of the symmetry breaking will be discussed in a latter section. $H_{\text{DMI}}$'s amplitude of 37 Oe is comparable to the results reported By Han *et al.* [17] but smaller than that in Ref. [18], possibly due to a much thicker Pt thickness weakening the long-range indirect exchange interaction. The observed phenomenon demonstrates beyond doubt that $M_{\text{PMA}}$ is subject to the effect of a sizable interlayer DMI coupling. Other frequently seen coupling effects, such as RKKY coupling, dipolar coupling [31] or symmetric interlayer exchange coupling unanimously fall into the category of symmetric exchange interaction, thus cannot explain the chiral nature of our findings. As for the antisymmetric intralayer DMI coupling scenario, we note that the two Pt layers in the wedged sample were specifically chosen to have identical thicknesses of 2.5 nm, alongside a low current density less than $1.32 \times 10^{10}$ A/m$^2$ to probe $R_H$, renders symmetry breaking contribution from damping-like spin-orbit torque (SOT) induced loop shift [21,32] minimal, as reported in single PMA layer structures [33]. This excludes the possibility of an intralayer DMI origin of the observed phenomenon. Lastly, we note that a reciprocal effect where the $M_{\text{IMA}}$ experiences an effective field in the form of $\boldsymbol{D} \times \boldsymbol{M}_{\text{PMA}}$ is verified with longitudinal magneto-optical Kerr effect during in-plane field sweep. This reciprocal effect manifests itself as a shift in the IMA hysteresis loop, similar to PMA layer's behavior (See Supplemental Material S1). From the applied in-plane field $|\boldsymbol{H}_\varphi|$ dependence of $H_{\text{offset}}$, we also rule out the possibility of a tilted easy axis effect scenario that could also give rise to a similar yet linear $M_{\text{IMA}}$-independent $H_{\text{offset}}$ [34,35] (See Supplemental Material S2).

Following the established protocol, we separately fabricated the second batch of samples with identical layer structure but alternatively has Hall bar devices which sequentially rotates 30° in the $xy$ plane to test the universality of the wedge-induced in-plane symmetry breaking. Devices with varying orientations are grown simultaneously onto a 1 cm × 1 cm substrate with serial number A to F (Fig. 2(a)). $H_{\text{offset}}$ is again measured, but here the six samples have independent $x_i y_i$ and $\varphi_{H,i}$ coordinates. The six data sets are presented in Fig. 2(b), with solid lines being sine fits to the data. One can observe the six sinusoidal fits not only have a similar magnitude of $H_{\text{DMI}} = 30 \pm 2$ Oe, their $\varphi_{H,i}$ dependence also shifts at a constant interval of approximately 30° from device A to F. We subsequently recalculate to convert $\varphi_{H,i}$ back into the $xy$ coordinate of the substrate (Fig. 2(a) lower left diagram) and plot the compilation of $H_{\text{DMI}}$ and $\varphi_D$ in Fig. 2(c), from which both $H_{\text{DMI}}$ and $\varphi_D$ are found to be universal among the six rotary devices. These results show that both the strength and the direction of the interlayer DMI are indeed universal throughout the whole substrate, in sharp contrast to results from Avci *et al.* with device-to-device variations [18].

Here we try to shed light on how a wedge-deposited structure can give rise to in-plane symmetry breaking effects. Dong-Soo *et al.* [17] previously attributed their symmetry breaking to a thickness gradient/lattice mismatch of their polycrystal multilayer, while Pacheco *et al.* [16] demonstrated that an external field induced anisotropy in the in-plane layer can be effective as well. Consequently, we first examine CoFeB's thickness gradient that could've occurred during the wedged deposition process. However, the thickness variation across a 20 μm-wide current channel is determined to be merely 5×10$^{-4}$ nm, much thinner than an atomic monolayer. While the thickness variation is miniscule, we argue that although our sputter-deposited CoFeB layer should be predominantly amorphous, different from Ref. [17], the oblique deposition approach can still generate nanocolumn structures with preferred orientation due to the shadow effect [17,36]. Such nanostructures can possibly further induce a structural in-plane symmetry breaking in the Pt mediating layer through a templating effect. We also note that both in-plane magnetic anisotropy and shadow effect induced by the wedge deposition may simultaneously exist in our samples (See Supplemental Material S3). This phenomenon is best demonstrated when comparing the two wedged CoFeB/Pt/Co/Pt samples used in our work. While both samples have their ***D*** directions decently perpendicular to the CoFeB atom flow direction (see Fig.1(a) and Fig.2(b) with $\varphi_D \approx 150°$ and $180°$, respectively) the $H_{DMI}$ strength and ***D*** direction still show discrepancies, possibly due to competition between the two symmetry breaking effects. Still, we emphasize that ***D*** in this work is universal across a sizable substrate.

After observing a universal behavior of the interlayer DMI on the same substrate, we propose a scenario utilizing interlayer DMI as a viable approach in achieving field-free current-driven magnetization switching. The symmetry of a PMA magnetization is normally preserved under current-induced SOT if the spin polarization direction is constrained in-plane without the usage of unconventional spin polarizations [37], therefore field-free switching cannot be realized. Prior to our work, numerous methods were proposed to break the mirror symmetry and to achieve a purely-current-driven magnetization switching, such as utilizing an intralayer DMI [38], tilted magnetic anisotropy [35,39], exchange bias [40-43] and epitaxial materials with broken mirror symmetry [37,44]. It is postulated here that when the IMA layer experiences magnetization switching, the interlayer DMI effective field will reverse its polarity as well, potentially assist in breaking the symmetry between up/down magnetization states or switching the PMA magnetization outright. Furthermore, the switching polarity of the PMA layer could be controlled by the relative orientation between $M_{IMA}$ and $D$, since the cross product of the two vectors dictates the orientation and strength of the DMI effective field. Assuming the $D$ vector points toward the $-x$ direction, the landscape of $H_{DMI}$ acting on $M_{PMA}$ is plotted in Fig. 3(a). When $M_{IMA}$'s angle, $\varphi_M$, lies within the first and second Quadrant (solid area), $H_{DMI}$ points toward $+z$ and maximizes at $\varphi_M = 90°$. On the contrary, $H_{DMI}$ points toward $-z$ and minimizes at $\varphi_M = 270°$ while $\varphi_M$ lies within the third and fourth Quadrant (dashed area). Utilizing rotary devices, two geometries with their respective current channel striking an angle of $45°$ ($135°$) with the $x$ axis are showcased in Fig. 3(b) left (right). For the $45°$ device, the applied pulse current ($I_{pulse}$) results in $M_{IMA}$ magnetization switching, its direction thus can take two values, $\varphi_M = 135°$ ($-45°$) when $I_{pulse}$ is positive (negative) and exceeds the critical switching value. In the case of $\varphi_M = 135°$ $M_{PMA}$ is pulled toward $+z$ due to a positive $H_{DMI}$ whilst $M_{PMA}$ tends to point toward $-z$ due to a negative $H_{DMI}$ when $\varphi_M = -45°$. A similar case is illustrated in the $135°$ device, but the $M_{PMA}$'s orientation dependence of the current polarity is the opposite to that of the $45°$ device.

The above model is confirmed by performing current-induced magnetization switching measurements on a magnet-free probe station, with current pulses of various amplitudes and pulse width of 50 ms. For both 45° and 135° devices, their perpendicular magnetization's deterministic reversal processes are shown in Fig. 3(c) and (d), respectively (blue data set). We observed that the two devices' switching polarities are not just opposite, but precisely follow the interlayer DMI effective field landscape that describes the $H_{\text{DMI}}$ response in relation to $\varphi_M$ (For in-plane field induced PMA switching, see Supplemental Material S4). Our model also implicitly suggests that $M_{\text{PMA}}$ and $M_{\text{IMA}}$ should exhibit coherent switching where the two magnetizations have identical switching currents, since the DMI effective field (~ 37 Oe) surpasses the $H_c$ (~ 25 Oe) of $M_{\text{PMA}}$. We thus perform consecutive current scans right after PMA switching in which $M_{\text{IMA}}$'s switching is probed by the UMR [45], a detection method that utilizes alternating $I_{\text{pulse}}$ to generate difference in magnetoresistance ($\Delta R$) (Fig.3 (c) and (d), red data set). From the UMR data, IMA's switching current in both 45° and 135° devices agrees relatively well with that of the PMA switching data, confirming coherent switching. We highlight that the IMA CoFeB's switching currents are slightly less than that of PMA Co's switching. When performing PMA Co switching experiments, it is mostly the Hall-cross (intersection) region that contributes to the Hall signal, while on the contrary the UMR detects signal that originated mainly from the Hall-bar (longitudinal) region. A previous work [46] pointed out the current density to be different in these two regions, with it being higher in the longitudinal region. This explains the lower switching current measured by UMR. Different from the PMA switching scheme, however, is that the switching polarity of the IMA layer remains unchanged between the two devices. It is because the in-plane magnetization switching is realized by the current-induced Oersted field and/or SOT originated from the adjacent Pt and thus the switching curve is solely governed by $I_{\text{pulse}}$'s polarity regardless of the direction of **D**.

We hereby comment that our devices can achieve deterministic and full switching. We highlight the switching process to be independent of magnetic history since both magnetizations are reinitialized simultaneously under applied current and thus always retain the exact polarity, unlike other field-free approaches that rely on exchange bias, orange peel effect [47,48] or even exotic spin currents [37]. The vanishingly small SOT acting on $\boldsymbol{M}_{\text{PMA}}$ also differentiates our work from the previously reported mechanism of the type-T scenario [49], in which simultaneous switching of the IMA and PMA layer was also demonstrated, though in that case the magnetic coupling is symmetric and the PMA switching requires sizable SOT, thus poses striking differences to our work. Finally, it is noted that the in-plane CoFeB layer can be replaced by a soft magnetic material, Permalloy (Py). All the characteristics of interlayer DMI coupling, from sizable $H_{\text{DMI}}$, universal **D** vector to field-free switching are retained (See Supplemental Material S5).

In conclusion, we demonstrated a sizable interlayer DMI within a magnetic multilayer system with orthogonal magnetic orientations. The existence of such an interlayer DMI is experimentally verified through an intricately calibrated field sweep procedure, revealing the chiral nature of the DMI with out-of-plane effective field strength $H_{\text{DMI}}$ = 37 Oe. Comparisons with control samples pinpoint interlayer DMI's origin to the oblique deposition induced in-plane symmetry breaking and thus extends the influence of DMI from 2D to 3D magnetic textures. Characterizations of devices with multiple orientations on the same substrate showcase the universal strength and direction of such interaction, beneficial to wafer-scale integration. Furthermore, all-electric (current-driven) 3D magnetic texture harnessing is demonstrated in the deterministic field-free magnetization switching experiment. These results unequivocally prove the potential of strong interlayer DMI to be employed when designing inter-connected 3D magnetic structures for field-free approaches within future spintronic devices.

## ACKNOWLEDGEMENTS

This work is supported by the Ministry of Science and Technology of Taiwan (MOST) under grant No. 111-2636-M-002-012 and by the Center of Atomic Initiative for New Materials (AI-Mat), National Taiwan University from the Featured Areas Research Center Program within the framework of the Higher Education Sprout Project by the Ministry of Education (MOE) in Taiwan under grant No. NTU-111L9008.

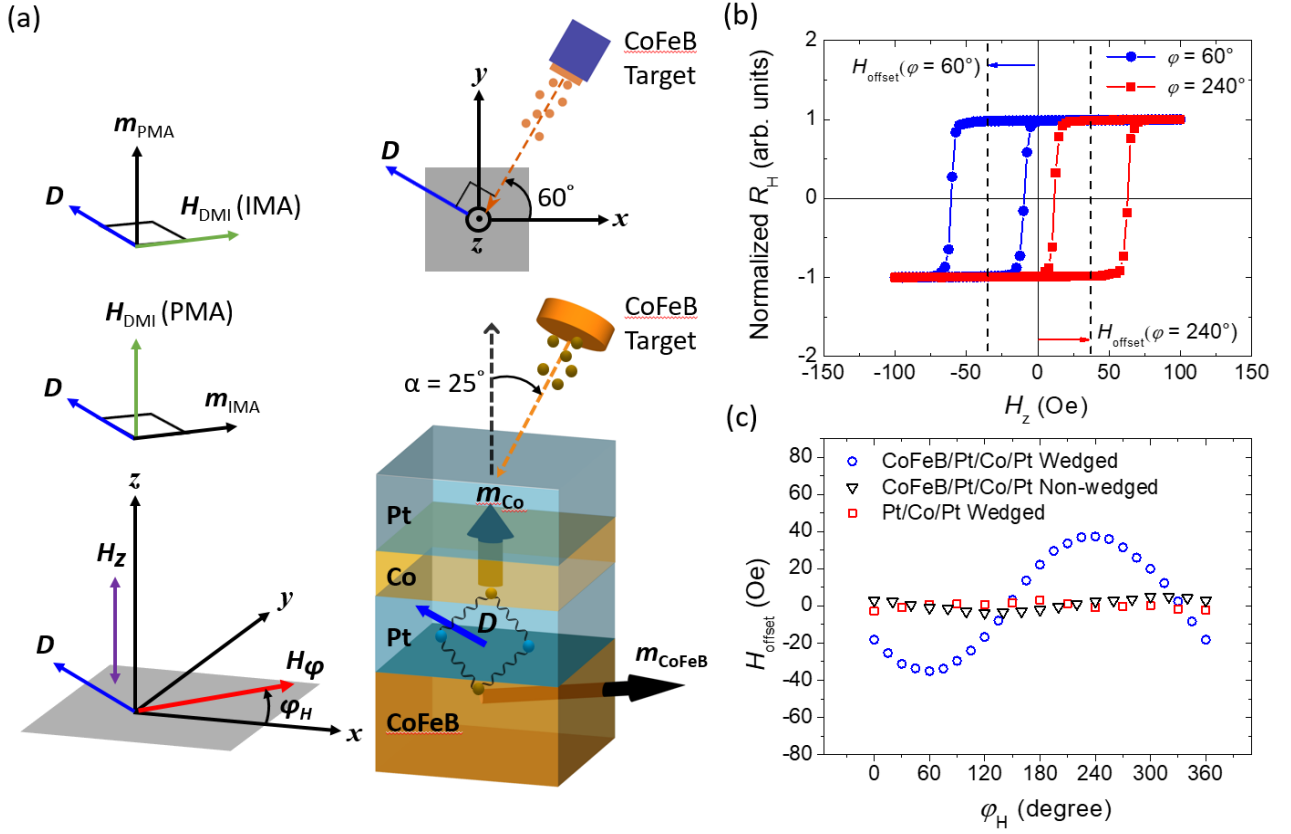

FIG. 1. Sketch of the magnetic multilayers, external fields, coordinate system and interlayer DMI's manifestation on $R_H$ hysteresis loops. (a) Illustration of the multilayer structure, with the CoFeB atom flow (target) tilted away from substrate normal at α=25°. Lower left diagram shows the coordinate system and the external field geometries. Upper left diagram demonstrates the cross-product relationship between $M_{IMA}$, $M_{PMA}$, $D$ and $H_{DMI}$ (PMA and IMA). Upper right diagram indicates the relation between the CoFeB atom flow (symmetry breaking vector) and the $D$ direction. (b) Representative Hall resistance loops. Field sweeps were performed by applying a constant field $H_\varphi$ in the $xy$ plane while sweeping $H_z$ field (see (a)). Blue and red data sets correspond to $\varphi_H = 60°$ and 240°, respectively with $|H_\varphi| = 300$ Oe. (c) $H_{offset}$ as a function of $\varphi_H$. The three samples used are the fully wedged CoFeB(2)/Pt(2.5)/Co(0.6)/Pt(2.5) sample and two control samples, namely fully

wedged Pt(2.5)/Co(0.6)/Pt(2.5) and non-wedged CoFeB(2)/Pt(2.5)/Co(0.6)/Pt(2.5).

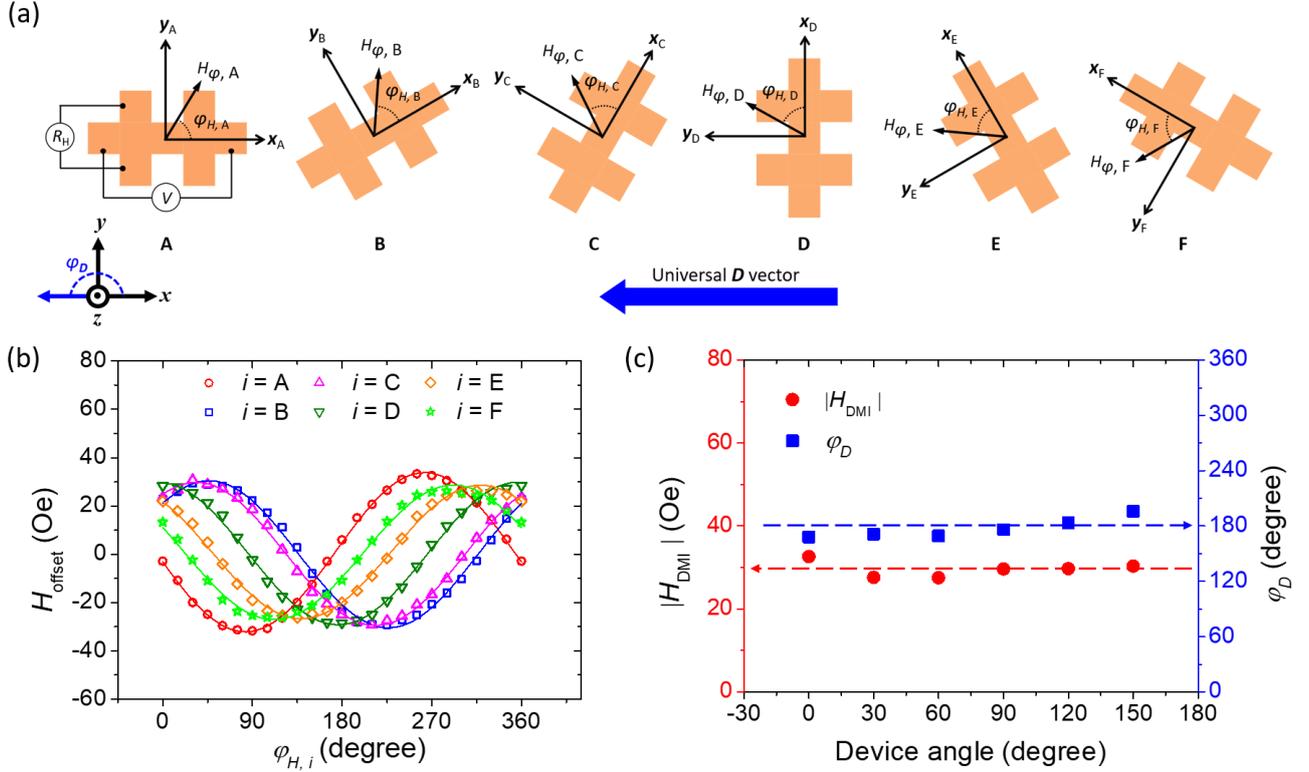

FIG. 2. Demonstration of universal interlayer DMI within the same substrate, among different devices. (a) Schematic of six rotating hall cross devices, namely device A to F. These six devices have their current channel direction rotated with an interval of 30°. Individual device has their independent $x_i y_i$ and $\varphi_{H,i}$ coordinates to differentiate their respective field sweeping procedures. (b) Results from interlayer DMI coupling induced hysteresis loop shift measurements, note the different coordinate systems for each individual device. Solid lines are sine fits to the data. $|\boldsymbol{H}_\varphi| = 300$ Oe throughout this set of experiments. (c) Compilation of $\varphi_D$ and $H_{DMI}$ extracted from (b) with sine fitting. The characteristic $\boldsymbol{D}$ vector's direction, $\varphi_D$, is recalculated to fit the universal $xy$ coordinate of the substrate.

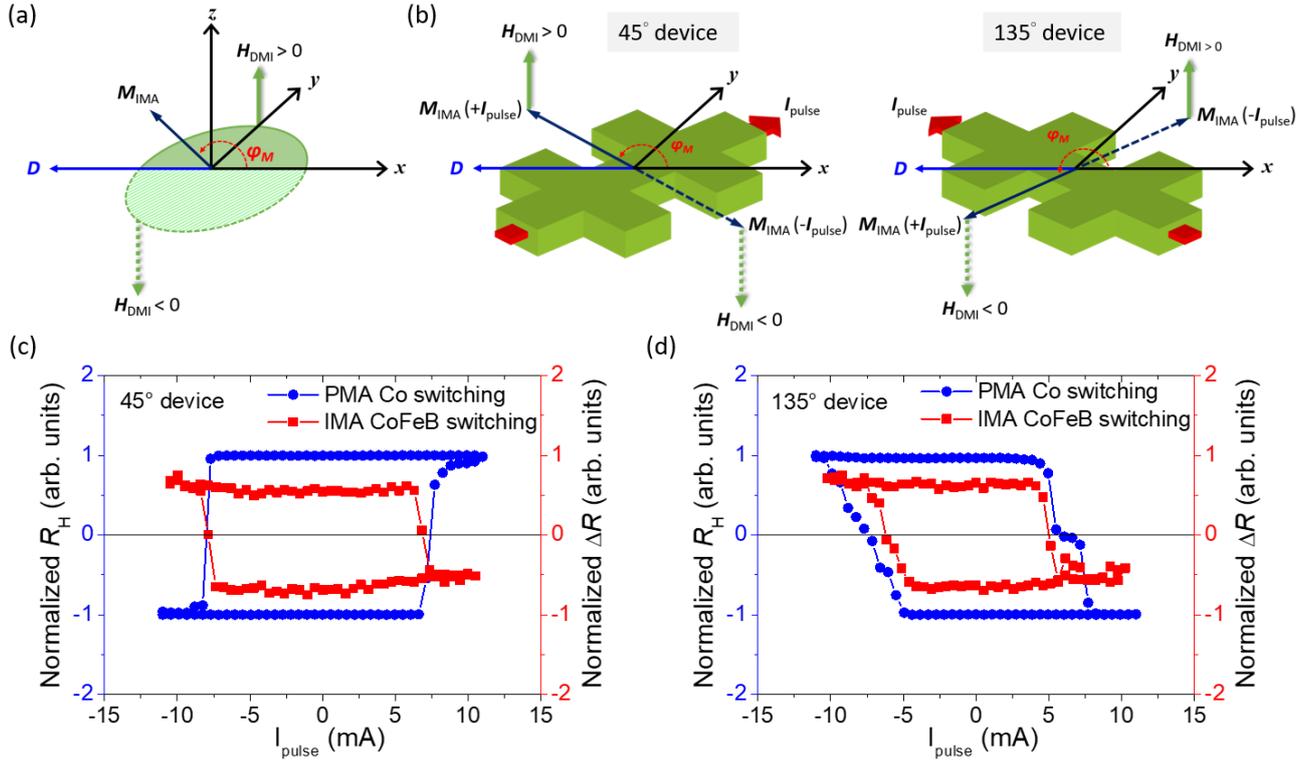

FIG. 3. Demonstration of current-driven field-free switching induced by the interlayer DMI. (a) Schematic of the relationships between $M_{IMA}$, $D$ vector and the resultant effective DMI field $H_{DMI}$. (b) Illustration of two devices on the same substrate, with their respective current channel directions strike 45° and 135° with $x$ axis, respectively. Following a conventional Oersted field and/or SOT switching scheme, $M_{IMA}$ can be switched between two states, prompting $H_{DMI}$ response with opposite chirality. (c) and (d) show current-induced magnetization switching of both PMA and IMA layers in a field-free configuration. PMA switching is sensed by $R_H$, while IMA switching is sensed by the UMR.

10854 (2016).

[44] L. Liu *et al.*, Symmetry-dependent field-free switching of perpendicular magnetization, Nat. Nanotechnol. **16**, 277 (2021).

[45] Y.-T. Liu, T.-Y. Chen, T.-H. Lo, T.-Y. Tsai, S.-Y. Yang, Y.-J. Chang, J.-H. Wei, and C.-F. Pai, Determination of Spin-Orbit-Torque Efficiencies in Heterostructures with In-Plane Magnetic Anisotropy, Phys. Rev. Appl. **13**, 044032 (2020).

[46] T.-Y. Tsai, T.-Y. Chen, C.-T. Wu, H.-I. Chan, and C.-F. Pai, Spin-orbit torque magnetometry by wide-field magneto-optical Kerr effect, Sci. Rep. **8**, 5613 (2018).

[47] N. Murray, W.-B. Liao, T.-C. Wang, L.-J. Chang, L.-Z. Tsai, T.-Y. Tsai, S.-F. Lee, and C.-F. Pai, Field-free spin-orbit torque switching through domain wall motion, Phys. Rev. B **100**, 104441 (2019).

[48] W. L. Yang *et al.*, Role of an in-plane ferromagnet in a T-type structure for field-free magnetization switching, Appl. Phys. Lett. **120** (2022).

[49] W. J. Kong *et al.*, Spin-orbit torque switching in a T-type magnetic configuration with current orthogonal to easy axes, Nat. Comm. **10**, 233 (2019).